\documentclass[aps,prd,amssymb,nofootinbib,twocolumn,epsf,showpacs]{revtex4}
\usepackage[usenames]{color} 
\usepackage{ulem} 
\usepackage{graphicx}
\usepackage{amssymb}
\usepackage{amsmath}
\usepackage{epstopdf}

\begin{document}

%%%%%%%%%%%%%%%%%%%%%%%%%%%%%%%%%%%%%%%%%%%%%%%%%%%%%%%%%%%%%%%%%%%%%%%%%%%%%%%
% A Spectral Approach to the Inverse Stellar Structure Problem
%%%%%%%%%%%%%%%%%%%%%%%%%%%%%%%%%%%%%%%%%%%%%%%%%%%%%%%%%%%%%%%%%%%%%%%%%%%%%%%

\title{Inverse Structure Problem for Neutron-Star Binaries}

\author{Lee Lindblom}

\affiliation{Center for Astrophysics and Space Sciences, University 
of California at San Diego, La Jolla, CA 92093, USA}

\date{\today}

\begin{abstract}
  Gravitational wave detectors in the LIGO/Virgo frequency band are
  able to measure the individual masses and the composite tidal
  deformabilities of neutron-star binary systems.  This paper
  demonstrates that high accuracy measurements of these quantities
  from an ensemble of binary systems can in principle be used to
  determine the high density neutron-star equation of state exactly.
  This analysis assumes that all neutron stars have the same
  thermodynamically stable equation of state, but does not use
  simplifying approximations for the composite tidal deformability or
  make additional assumptions about the high density equation of
  state.
\end{abstract}


\maketitle

%%%%%%%%%%%%%%%%%%%%%%%%%%%%%%%%%%%%%%%%%%%%%%%%%%%%%%%%%%%%%%%%%%%%%%%%%%%%%%%

\section{Introduction}
\label{s:Introduction}

The masses, $M$, and the tidal deformabilities, $\Lambda$, of neutron
stars can (in principle) be measured by observations of the
gravitational waves emitted during the last stages of the inspiral of
neutron-star binary systems~\cite{Hinderer2008a}.
%\cite{Hinderer2008, Hinderer2009,
%  Read:2009yp, Hinderer2010, Lackey2012, Damour2012, Bernuzzi2012,
%  Read2013, Lackey2013, Pozzo2013, Maselli2013}.)
Since all neutron stars are expected to have the same equation of
state, accurate measurements of $M$ and $\Lambda$ for an ensemble of
neutron stars could be used to determine the high density portion of
the neutron star equation of state exactly by solving the inverse
stellar structure problem~\cite{Lindblom2014, Lindblom2016,
  Lindblom2014a}.

Unfortunately, the individual tidal deformabilities of the stars in a
neutron-star binary system are not accurately observable by
gravitational wave detectors operating in the LIGO/Virgo frequency
band.\footnote{Tidal distortion effects first appear in the
  post-Newtonian expansion of the gravitational waveform at order
  $(v/c)^{10}$ as a term proportional to a composite deformability
  parameter.  It is only at even higher order that additional
  terms appear that would allow the deformabilities of the individual
  stars to be determined.  Gravitational wave detectors operating in
  the LIGO/Virgo frequency band are never likely to be able to measure
  those high order terms in neutron-star binary systems.}  Instead a
composite tidal deformability $\tilde\Lambda$, representing the
deformability of the binary system as a whole, is observable with such
detectors.  This composite tidal deformability is related to the
properties of the individual stars by
\begin{eqnarray}
  \tilde\Lambda &=& \frac{16}{13}\frac{M_1^4(M_1+12 M_2)\Lambda_1
    +M_2^4(M_2+12M_1)\Lambda_2}{(M_1+M_2)^5},\quad
  \label{e:tildeLambda}
\end{eqnarray}
where $\Lambda_1$ and $\Lambda_2$ are the tidal deformabilities, and
$M_1\ge M_2$ are the masses of the individual
stars~\cite{Hinderer2008a, Hinderer2008, Hinderer2009}.  The
observation of gravitational waves from a neutron-star binary,
GW170817, provides the first (and at present only) observation of
$M_1$, $M_2$ and $\tilde\Lambda$ for a binary
system~\cite{GW170817a,GW170817b}.

The purpose of this paper is to explore the extent to which
measurements of the masses, $M_1$ and $M_2$, and the composite tidal
deformabilities, $\tilde\Lambda$, of neutron-star binaries can in
principle be used to determine the high density portion of the
neutron-star equation of state.  Could such measurements determine the
equation of state exactly (assuming the measurement errors could be
made arbitrarily small) through the solution of some appropriate
inverse structure problem?  Or, are such measurements only able to
constrain the equation of state in some way?

An inverse structure problem determines the equation of state of the
matter in an astrophysical system using measurements of the
macroscopic properties of that system.  Mathematically well posed
inverse structure problems do exist for individual neutron
stars~\cite{Lindblom1992, Lindblom2012, Lindblom2014, Lindblom2016,
  Lindblom2014a}.  In particular, given a complete knowledge of the
curve of observables, $M(p_c)$ and $\Lambda(p_c)$ (parameterized for
example by the central pressures $p_c$ of the stars), this curve
exactly determines the equation of state, $\epsilon=\epsilon(p)$, a
curve in the energy density $\epsilon$, pressure $p$ space.  It is not
surprising that the stellar structure equations determine this unique
relationship (and inverse relationship) between these curves.  It is
less obvious that an analogous inverse structure problem exists for
binary systems.  Does a complete knowledge of the two-dimensional
surface of observables for binary systems, $M_1(p_{1c})$,
$M_2(p_{2c})$ and $\tilde\Lambda(p_{1c},p_{2c})$ (parameterized for
example by the central pressures, $p_{1c}$ and $p_{2c}$, of each star)
determine the equation of state exactly as well?

The inverse structure problem for neutron-star binaries does have an
almost trivial formal solution.  Given a complete knowledge of the
surface of observables, $\left\{M_1(p_{1c}), M_2(p_{2c}),
\tilde\Lambda(p_{1c},p_{2c})\right\}$, the equation of state can be
determined exactly by restricting attention to equal-mass binaries:
$M_1(p_{1c})=M_2(p_{2c})$, so that $p_{1c}=p_{2c}$ and
$\Lambda_1(p_{1c})=\Lambda_2(p_{1c})=\tilde\Lambda(p_{1c},p_{1c})$.
The inverse structure problem for binaries in this special case
reduces to the single neutron-star inverse structure problem, and that
problem can be solved exactly in various ways~\cite{Lindblom2014,
  Lindblom2016, Lindblom2014a}.  Unfortunately, observations of
precisely equal mass binary systems will never be available.  So, the
interesting question is not whether the inverse structure problem for
binaries has a formal solution, but rather how (and how well) it can
be solved using measurements from a random ensemble of unequal mass
binary systems.

The method proposed here for solving the inverse structure problem for
binaries is a fairly straightforward generalization of the method
developed previously for individual neutron stars~\cite{Lindblom2014,
  Lindblom2016, Lindblom2014a}.  Consider a random ensemble of data
points, $\left\{M_{1i}, M_{2i}, \tilde\Lambda_i\right\}$ for
$i=1,...,N_B$, taken from the exact surface of observables.  The goal
is to find an equation of state whose model observables match these
data.  This is done by introducing a parametric representation of the
equation of state, $\epsilon=\epsilon(p,\gamma_k)$, where the
$\gamma_k$ are parameters whose values can be adjusted to approximate
any equation of state to any desired accuracy~\cite{Read:2008iy,
  Lindblom2010, Lindblom2018}.  Given this equation of state model,
and choices for the central pressures of each of the stars in the
binary, $p^i_{1c}$ and $p^i_{2c}$, it is straightforward to integrate
the stellar structure equations to determine the masses
$M_1(p^i_{1c},\gamma_k)$ and $M_2(p^i_{2c},\gamma_k)$, and the tidal
deformabilities $\Lambda_1(p^i_{1c},\gamma_k)$ and
$\Lambda_2(p^i_{2c},\gamma_k)$. The resulting model observables
$M_1(p^i_{1c},\gamma_k)$, $M_2(p^i_{2c},\gamma_k)$ and
$\tilde\Lambda(p^i_{1c},p^i_{2c},\gamma_k)$ from
Eq.~(\ref{e:tildeLambda}), are then compared to the exact data using
the quantity $\chi^2$ that measures the modeling error:
\begin{eqnarray}
  &&\!\!\!\!\!
  \chi^2(p_{1c}^i,p_{2c}^i,\gamma_k)=\nonumber\\
&&\qquad\frac{1}{N_\mathrm{B}}\sum_{i=1}^{N_\mathrm{B}}\left\{
  \left[\log\left(\frac{M_1(p_{1c}^i,\gamma_k)}{M_{1i}}\right)\right]^2\right
  .\nonumber\\
  &&\qquad\qquad\qquad\quad
  +
  \left.\left[\log\left(\frac{M_2(p_{2c}^i,\gamma_k)}
    {M_{2i}}\right)\right]^2\right
  .\nonumber\\
&&\qquad\qquad\qquad\quad
+\left.\left[\log\left(\frac{\tilde\Lambda(p_{1c}^i,p_{2c}^i,\gamma_k)}
{\tilde\Lambda_i}\right)\right]^2\right\}.
\qquad
\label{e:PBasedChiSquareLMDef}
\end{eqnarray}
The error measure, $\chi^2$, is then minimized over the
$2N_B+N_\gamma$ dimensional space of parameters
$\left\{p^i_{1c},p^i_{2c},\gamma_k\right\}$.  The location of this
minimum determines an equation of state model,
$\epsilon=\epsilon(p,\gamma_k)$, whose stellar models best fit the
observations.

The equation of state, $\epsilon=\epsilon(p,\gamma_k)$, obtained by
minimizing $\chi^2$ in Eq.~(\ref{e:PBasedChiSquareLMDef}) provides an
approximation to the physical neutron-star equation of state.  If this
method of solving the inverse structure problem for binaries is
successful, these approximate equations of state should become more
accurate as $N_\gamma$ the number of parameters in the equation of
state model, and as $N_B$ the number of binary data points are
increased.

The remainder of this paper describes a series of numerical tests that
illustrate how well this inversion method actually works in practice.
Section~\ref{s:MockBinaryData} describes the construction of mock
data, $\left\{M_{1i}, M_{2i}, \tilde\Lambda_i\right\}$ for
$i=1,...,N_B$, from a known equation of state.
Section~\ref{s:ParametricRepresentations} describes the parametric
representations of the equation of state used in these tests.  These
representations, based on spectral expansions of the adiabatic index,
are shown to converge exponentially to the ``exact'' equation of state
used for the mock data in Sec.~\ref{s:MockBinaryData}.
Section~\ref{s:NumericalInversionTests} solves the inverse structure
problem with these mock binary data using the method described above
to determine approximate parametric model equations of state.  The
accuracy of these model equations of state are then evaluated by
comparing them to the original ``exact'' equation of state used to
construct the mock data.  These results are described at length in
Secs.~\ref{s:NumericalInversionTests} and \ref{s:Discussion}.  In
summary: the errors in the equation of state models decrease
exponentially in these tests as the number of parameters $N_\gamma$ is
increased.  This method for solving the inverse structure problem for
binaries therefore works very well.

%%%%%%%%%%%%%%%%%%%%%%%%%%%%%%%%%%%%%%%%%%%%%%%%%%%%%%%%%%%%%%%%%%%%%%%%%%%%%%%

\section{Mock Binary Data}
\label{s:MockBinaryData}

%%%%%%%%%%%%%%%%%%%%%%%%%%%%%%%%%%%%%%%%%%%%%%%%%%%%%%%%%%%%%%%%%%%%%%%%%%%%%%%

Gravitational wave observations of neutron-star binaries can measure
the masses, $M_1$ and $M_2$, and the composite tidal deformabilities
$\tilde\Lambda$ of those systems.  Mock data,
$\left\{M_{1i},M_{2i},\tilde\Lambda_i\right\}$ for $i=1,...,N_B$, are
constructed in this section, to be used in
Sec.~\ref{s:NumericalInversionTests} to test the solution to the
inverse structure problem for binaries outlined in
Sec.~\ref{s:Introduction}.  These mock data are constructed from the
simple pseudo-polytrope,
\begin{equation}
  p=p_0\left(\frac{\epsilon}{\epsilon_0}\right)^2,
  \label{e:polytropiceos}
\end{equation}
chosen as the exemplar ``exact'' equation of state in part because its
adiabatic index is similar to more realistic models of neutron-star
matter.  For these tests the constants $p_0$ and $\epsilon_0$ are
chosen to have the values $p_0=8\times 10^{33}$ and
$\epsilon_0=2\times 10^{14}$ in cgs units.  The resulting equation of state
produces a maximum mass neutron-star model of about $2.339M_\odot$.

The goal of the numerical tests performed in
Sec.~\ref{s:NumericalInversionTests} is to determine how well and how
accurately the method for solving the inverse structure problem
described in Sec.~\ref{s:Introduction} actually works.  To do this
effectively, extremely accurate mock data are needed.  The stellar
structure equations can be solved numerically more accurately using an
enthalpy based rather than the standard pressure based form of those
equations~\cite{Lindblom1992}.\footnote{The enthalpy of the star
  approaches zero linearly at the surface of the star, while the
  pressure approaches zero as a relatively high power of the distance
  from the surface.  Consequently it is much more difficult to
  determine the location of the surface (and the other macroscopic
  observables of the star) accurately using the standard pressure
  based forms of the equations.}  Consequently it is useful to
re-write the equation of state in terms of the enthalpy $h$.  The
simple equation of state used for these tests,
Eq.~(\ref{e:polytropiceos}), has the following enthalpy based form,
\begin{eqnarray}
  \epsilon(h)&=&\frac{\epsilon_0^2 c^2}{p_0}\left(e^{h/2}-1\right),
  \label{e:enthalpyeos1}\\
  p(h)&=&\frac{\epsilon_0^2 c^4}{p_0}\left(e^{h/2}-1\right)^2.
  \label{e:enthalpyeos2}
\end{eqnarray}

\begin{figure}[!htb]
%\centerline{\includegraphics[width=3in]{Binaries.eps}}
\centerline{\includegraphics[width=3in]{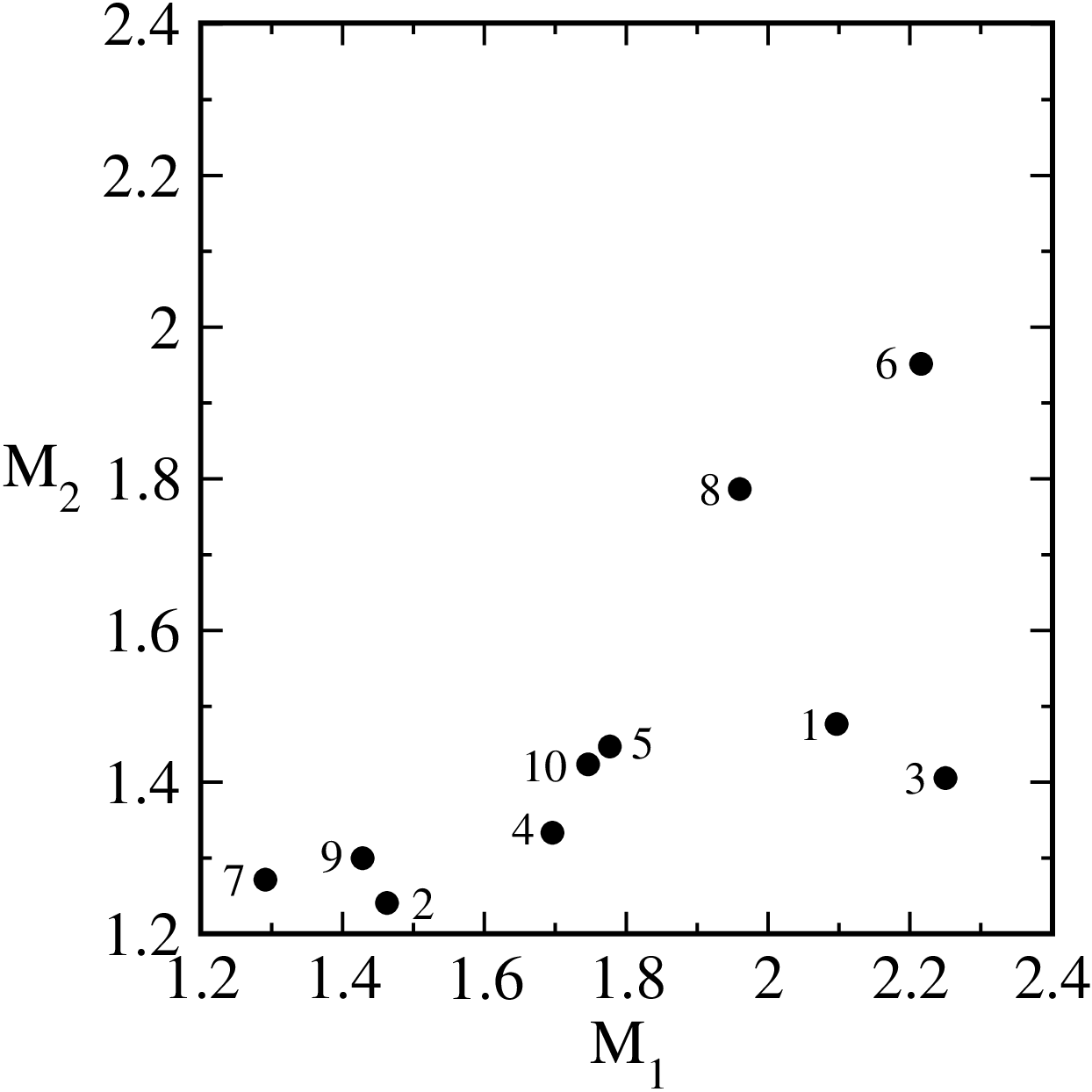}}
\caption{\label{f:Binaries} Points indicate the randomly chosen mass
  pairs $M_1\geq M_2$ included in the mock data set.}
\end{figure}

The masses $M_1$ and $M_2$ in these mock data are computed by solving
the standard Oppenheimer-Volkoff equations~\cite{Oppenheimer1939}
transformed into enthalpy based forms~\cite{Lindblom1992}. And, the
tidal deformabilities are computed using the equations derived by
Hinderer~\cite{Hinderer2008, Hinderer2009}, but transformed into
enthalpy based forms~\cite{Lindblom2014, Lindblom2016}.  The central
enthalpies, $h^i_{1c}$ and $h^i_{2c}$, for the stars in each mock
binary system are chosen with a random number generator from the range
needed to produce stars with masses between $1.2M_\odot$ and the
maximum mass $2.339M_\odot$.\footnote{The stellar models used for the
  mock data were constructed in a two step process.  First a large
  collection of $N_\mathrm{stars}$ models were constructed whose
  central enthalpies are given by $h_c^n= h_\mathrm{min}+
  (h_\mathrm{max}-h_\mathrm{min}) (n/N_\mathrm{stars})^2$ for
  $n=1,...,N_\mathrm{stars}$, with $h_\mathrm{min}$ and
  $h_\mathrm{max}$ being the central enthalpies of the models with
  $M=1.2M_\odot$ and $M=2.339M_\odot$ respectively.  This choice of
  the $h_c^n$ produces a collection of stellar models
  $\{M_n,\Lambda_n\}$ having (roughly) equally spaced masses.  The
  second step uses a random number generator, ran2 from
  Ref.~\cite{numrec_f}, to generate a uniformly distributed random
  sequence of integers $1\leq \ell \leq N_\mathrm{stars}=1000$.  This
  random sequence of integers is then used to select the particular
  stellar models used as the mock data for these tests,
  $\{M_i,\Lambda_i\}$, from the much larger collection of models
  $\{M_n,\Lambda_n\}$. } Figure~\ref{f:Binaries} illustrates the
resulting mock binary systems that are used in the numerical tests in
Sec.~\ref{s:NumericalInversionTests}.  The number labels of the
mass-pair points indicate the (randomly chosen) order in which the
models are used in the inversion tests.  For example, a test involving
$N_B$ binaries would use the data points labeled $1,...,N_B$.
%

%%%%%%%%%%%%%%%%%%%%%%%%%%%%%%%%%%%%%%%%%%%%%%%%%%%%%%%%%%%%%%%%%%%%%%%%%%%%%%%
\section{Parametric Representations of the Equation of State}
\label{s:ParametricRepresentations}

%%%%%%%%%%%%%%%%%%%%%%%%%%%%%%%%%%%%%%%%%%%%%%%%%%%%%%%%%%%%%%%%%%%%%%%%%%%%%%%

This section describes the parametric representations of the equation
of state used in the numerical tests of the inverse structure problem
in Sec.~\ref{s:NumericalInversionTests}.  Since these tests use
enthalpy based representations of the stellar structure equations,
enthalpy based parametric representations of the equation of state are
needed.  The most efficient representations of this type presently
available are based on spectral representations of the adiabatic index
$\Gamma(h)$~\cite{Lindblom2010}.  The best studied example uses the
expansion,
\begin{equation}
  \log \Gamma(h,\gamma_k)=\sum_{k=1}^{N_\gamma}\gamma_k
  \left[\log\left(\frac{h}{h_0}\right)\right]^{k-1},
  \label{e:spectralgamma}
\end{equation}
where the $\gamma_k$ are adjustable parameters, and $h_0$ determines
the low density limit of the domain where the spectral representation
is to be used.  For these tests the constant $h_0$ is chosen to
correspond to a density at the outer boundary of the neutron-star core
$\epsilon_0=\epsilon(h_0)=2\times 10^{14}$ g/cm${}^3$.  Below this
density the equation of state is assumed to be known, and is taken in
our tests to be the exact equation of state given in
Eqs.~(\ref{e:enthalpyeos1}) and (\ref{e:enthalpyeos2}).  Given this
expression for $\Gamma(h,\gamma_k)$, the parametric equation of state
itself is determined by the expressions~\cite{Lindblom2010}
\begin{eqnarray}
  p(h,\gamma_k)&=&
  p_0 \exp\left[\int_{h_0}^h \frac{e^{h'}dh'}{\mu(h',\gamma_k)}\right],
  \label{e:PressueH}\\
  \epsilon(h,\gamma_k)&=& p(h,\gamma_k)
  \frac{e^h -\mu(h,\gamma_k)}{\mu(h,\gamma_k)},
\label{e:EnthalpyH}
\end{eqnarray}
where $\mu(h,\gamma_k)$ is defined as,
\begin{eqnarray}
\mu(h,\gamma_k) = \frac{p_0\, e^{h_0}}{\epsilon_0 + p_0} 
+ \int_{h_0}^h \frac{\Gamma(h',\gamma_k)-1}{\Gamma(h',\gamma_k)} e^{h'}dh'.
\label{e:TildeMuDef}
\end{eqnarray}

\begin{figure}[!htb]
%\centerline{\includegraphics[width=3in]{Delta_n.eps}}
\centerline{\includegraphics[width=3in]{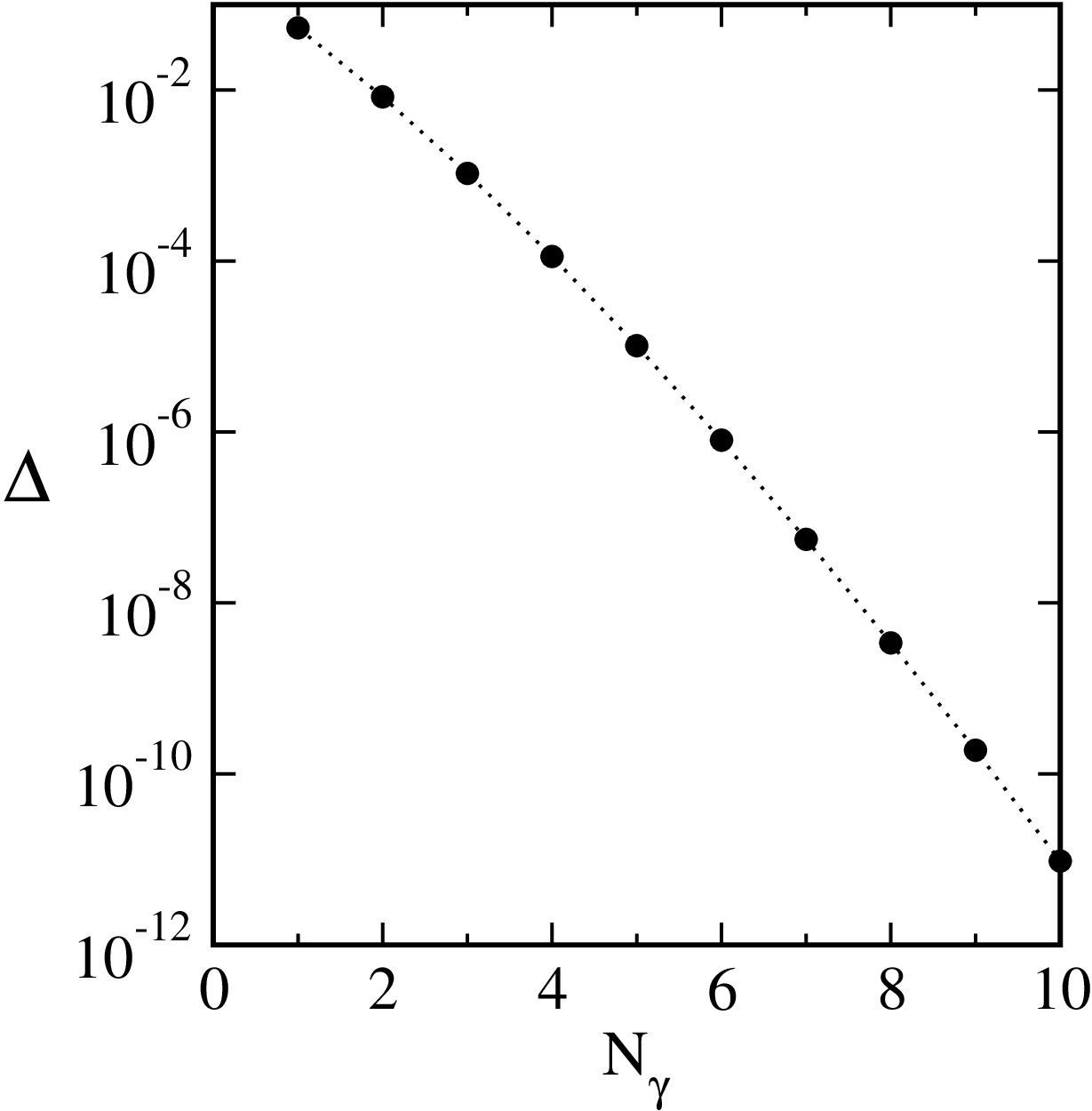}}
\caption{\label{f:Delta_n} Points illustrate the average errors
  $\Delta$ of the enthalpy-based spectral representations of the
  ``exact'' equation of state as a function of the order of the
  spectral representation, $N_\gamma$.}
\end{figure}

These parametric equations of states have been used successfully to
represent a variety of realistic nuclear-theory model equations of
state, with errors that converge toward zero as the number of
parameters $N_\gamma$ is increased~\cite{Lindblom2010, Lindblom2018}.
These representations are used in Sec.~\ref{s:NumericalInversionTests}
as approximations to the ``exact'' equation of state as determined by
the mock binary data from Sec.~\ref{s:MockBinaryData}.  It is useful
to understand, therefore, how well these parametric representations
are able to represent this ''exact'' equation of state.  The adiabatic
index for the ``exact'' equation of state of
Eq.~(\ref{e:polytropiceos}) is given by
\begin{equation}
  \Gamma(h)=\frac{\epsilon\, c^2 + p}{p\,c^2}\,\frac{dp}{d\epsilon}=
  2\frac{\epsilon\, c^2+p}{\epsilon\, c^2} =2+2\left(1-e^{h/2}\right)^2.
\end{equation}
While this $\Gamma(h)$ is quite simple, its representation as the
spectral expansion given in Eq.~(\ref{e:spectralgamma}) requires an
infinite number of terms.  The optimal values of the parameters
$\gamma_k$ can be estimated by minimizing the equation of state error
measure $\Delta(\gamma_k)$, defined as
\begin{equation}
\Delta^2(\gamma_k)=\frac{1}{N}
\sum_{i=1}^N\left[\log\left(\frac{\epsilon(h_i,\gamma_k)}{\epsilon_i}
  \right)\right]^2
\label{e:Deltadef}
\end{equation}
with respect to the $N_\gamma$ spectral parameters $\gamma_k$.  The
sum in this expression is taken over $N\approx 85$ points taken from
an exact equation of state table, equally spaced in $\log\epsilon$ in
the density range $2\times 10^{14} \leq \epsilon_i\leq 1.8895\times
10^{15}$ g/cm${}^3$ that covers the high density cores of all neutron
stars with this equation of state. This sum measures the differences
between the parametric equation of state densities
$\epsilon(h_i,\gamma_k)$ with $N$ exact densities
$\epsilon_i=\epsilon(h_i)$.  Figure~\ref{f:Delta_n} shows the minimum
values of $\Delta$ as a function of the number of spectral parameters
$N_\gamma$.  These parametric representations therefore converge
exponentially toward Eq.~(\ref{e:polytropiceos}), and
Fig.~\ref{f:Delta_n} provides a best-case estimate of the accuracy
that the approximate solutions to the inverse problem in
Sec.~\ref{s:NumericalInversionTests} might achieve.

Based on our understanding of other spectral representations, like
Fourier series, the spectral parametric representations used here are
expected to converge exponentially for all smooth equations of state.
The rate of exponential convergence will depend, however, on the
detailed structure of the particular equation of state.  Equations of
state having more ``structure'' than the simple pseudo-polytrope
studied here will converge more slowly.  Spectral parametric
representations of equations of state having phase transitions
(i.e. discontinuities in the equation of state or its derivatives) are
also expected to converge, however the rate of convergence in those
cases are expected to be polynomial rather than exponential.

%%%%%%%%%%%%%%%%%%%%%%%%%%%%%%%%%%%%%%%%%%%%%%%%%%%%%%%%%%%%%%%%%%%%%%%%%%%%%%%
\section{Numerical Inversion Tests}
\label{s:NumericalInversionTests}

%%%%%%%%%%%%%%%%%%%%%%%%%%%%%%%%%%%%%%%%%%%%%%%%%%%%%%%%%%%%%%%%%%%%%%%%%%%%%%%

The goal of the inverse structure problem for binaries is to determine
the equation of state from a knowledge of the observables
$\left\{M_1(h_{1c}),M_2(h_{2c}), \tilde\Lambda(h_{1c},h_{2c})\right\}$
(parameterized here by the central enthalpies $h_{1c}$ and $h_{2c}$ of
each star).  Let $\left\{M_{1i},M_{2i},\tilde\Lambda_i\right\}$ for
$i=1,...,N_B$ denote a random ensemble of points from the exact
surface of observables, and let $\epsilon=\epsilon(h,\gamma_k)$ and
$p=p(h,\gamma_k)$ denote a family of parametric equations of state.
The proposal is to construct approximate solutions to this inverse
structure problem by minimizing the difference between models of the
observables $\left\{M_1(h_{1c},\gamma_k),M_2(h_{2c},\gamma_k),
\tilde\Lambda(h_{1c},h_{2c},\gamma_k)\right\}$ based on the parametric
equation of state, and the observational data points
$\left\{M_{1i},M_{2i},\tilde\Lambda_i\right\}$.  This difference is
measured using the modeling error measure
$\chi^2(h_{1c}^i,h_{2c}^i,\gamma_k)$, defined by
\begin{eqnarray}
  &&\!\!\!\!\!
  \chi^2(h_{1c}^i,h_{2c}^i,\gamma_k)=\nonumber\\
&&\qquad\frac{1}{N_\mathrm{B}}\sum_{i=1}^{N_\mathrm{B}}\left\{
  \left[\log\left(\frac{M_1(h_{1c}^i,\gamma_k)}{M_{1i}}\right)\right]^2\right
  .\nonumber\\
  &&\qquad\qquad\qquad\quad
  +  \left.\left[\log\left(\frac{M_2(h_{2c}^i,\gamma_k)}
    {M_{2i}}\right)\right]^2\right
  .\nonumber\\
&&\qquad\qquad\qquad\quad
+\left.\left[\log\left(\frac{\tilde\Lambda(h_{1c}^i,h_{2c}^i,\gamma_k)}
{\tilde\Lambda_i}\right)\right]^2\right\}.
\qquad 
\label{e:ChiSquareLMDef}
\end{eqnarray}
The best-fit model is identified by minimizing the modeling error
$\chi^2(h_{1c}^i,h_{2c}^i,\gamma_k)$ with respect to the
$2N_B+N_\gamma$ parameters $\left\{h_{1c}^i,h_{2c}^i,
\gamma_k\right\}$.  The parametric equation of state
$\epsilon=\epsilon(h,\gamma_k)$ and $p=p(h,\gamma_k)$ with $\gamma_k$
evaluated at this minimum is an approximate solution to the inverse
structure problem.

The most difficult step in this approach is finding the minimum of
$\chi^2(h_{1c}^i,h_{2c}^i,\gamma_k)$ numerically.  The minimization
method used for these tests is the Levenberg-Marquardt
algorithm~\cite{numrec_f}.  This is a steepest descent type algorithm
that requires as input the value of the function,
$\chi^2(h_{1c}^i,h_{2c}^i,\gamma_k)$, and its partial derivatives with
respect to each of the parameters.  The needed partial derivatives can
be constructed from $\partial M/\partial h_c$, $\partial
\Lambda/\partial h_c$, $\partial M/\partial\gamma_k$ and
$\partial\Lambda/\partial \gamma_k$ (computed for these tests using
the methods described in Refs.~\cite{Lindblom2014, Lindblom2016}) plus
the derivatives
\begin{eqnarray}
  \frac{\partial\tilde\Lambda}{\partial M_1}&=&
  -\frac{16M_1^3M_2(7M_1-48M_2)\Lambda_1}{13(M_1+M_2)^6}\nonumber\\
&&   \qquad-\frac{16M_2^4(48M_1-7M_2)\Lambda_2}{13(M_1+M_2)^6},\\
  \frac{\partial\tilde\Lambda}{\partial M_2}&=&
  \frac{16M_1^4(7M_1-48M_2)\Lambda_1}{13(M_1+M_2)^6}\nonumber\\
&&   \qquad+\frac{16M_1M_2^3(48M_1-7M_2)\Lambda_2}{13(M_1+M_2)^6},\\
  \frac{\partial\tilde\Lambda}{\partial \Lambda_1}&=&
  \frac{16 M_1^4(M_1+12M_2)}{13(M_1+M_2)^5},\\
  \frac{\partial\tilde\Lambda}{\partial \Lambda_2}&=&
  \frac{16 M_2^4(M_2+12M_1)}{13(M_1+M_2)^5}.
\end{eqnarray}

The Levenberg-Marquardt minimization method is very fast and very
accurate at locating the local minimum close to any given initial
parameter point.  It often fails to find the smallest minimum,
however, if the function has many local minima.  To avoid unwanted
local minima, and to speed up the calculation, the numerical
minimizations performed for these tests were initialized using the
exact values of the parameters $h_{1c}^i$ and $h_{2c}^i$ from
Sec.~\ref{s:MockBinaryData}, and the best-fit values of the parameters
$\gamma_k$ described in Sec.~\ref{s:ParametricRepresentations}.  The
minimization procedure is iterated as many times as needed (typically
less than ten) until $\chi$ is unchanged from one step to the
next.\footnote{In the analysis of real neutron-star observations, it
  will not be possible to know a priori what the optimal parameters
  $h_{1c}^i$, $h_{2c}^i$ and $\gamma_k$ are likely to be.  In this
  case it will almost certainly be necessary to adopt more powerful
  computational methods for locating the absolute minimum of the
  complicated non-linear function $\chi^2(h_{1c}^i,
  h_{2c}^i,\gamma_k)$.}

\begin{figure}[!ht]
%\centerline{\includegraphics[width=3in]{chisqr_data.eps}}
\centerline{\includegraphics[width=3in]{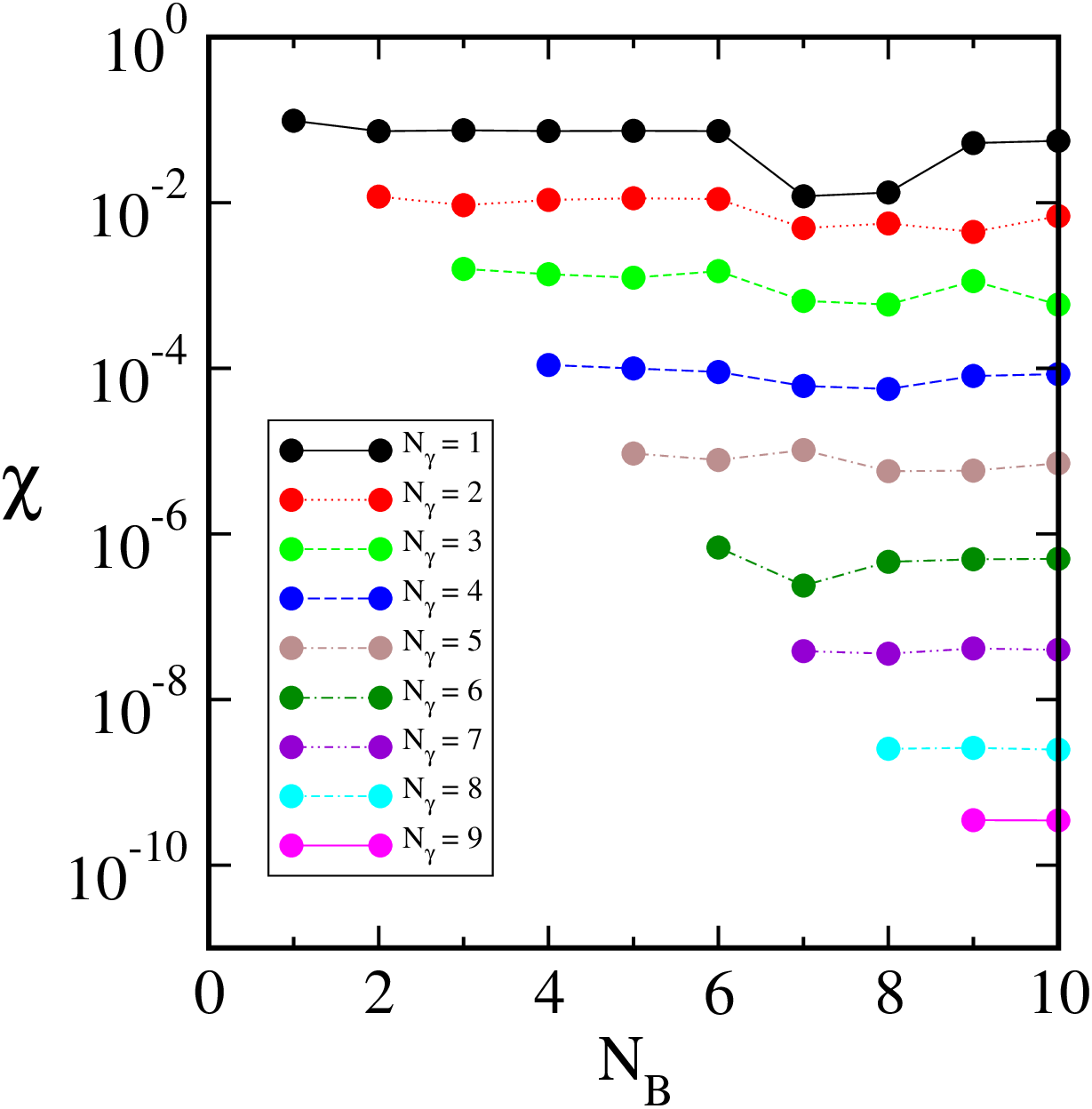}}
\caption{\label{f:chisqr_data} Curves indicate the minimum values of
  $\chi(h_{1c}^i,h_{2c}^i,\gamma_k)$ achieved for different numbers
  $N_\gamma$ of spectral parameters, and different numbers $N_B$ of
  mock binary data points.}
\end{figure}

Figure~\ref{f:chisqr_data} illustrates the minimum values of $\chi$
obtained in this way for different values of $N_\gamma$ and $N_B$.
The equations used to locate the minimum of $\chi$ are degenerate
whenever the number of parameters, $2N_B+N_\gamma$, is less than the
number of data points, $3N_B$.  Consequently these minima were only
computed for $N_\gamma\leq N_B$. This figure shows that the
numerically determined values of the minima of $\chi$ decrease
exponentially as the number of equation of state parameters $N_\gamma$
is increased.  These minima are relatively insensitive to the values
of $N_B$ for fixed values of $N_\gamma$.

\begin{figure}[!ht]
%\centerline{\includegraphics[width=3in]{chisqr_eos.eps}}
\centerline{\includegraphics[width=3in]{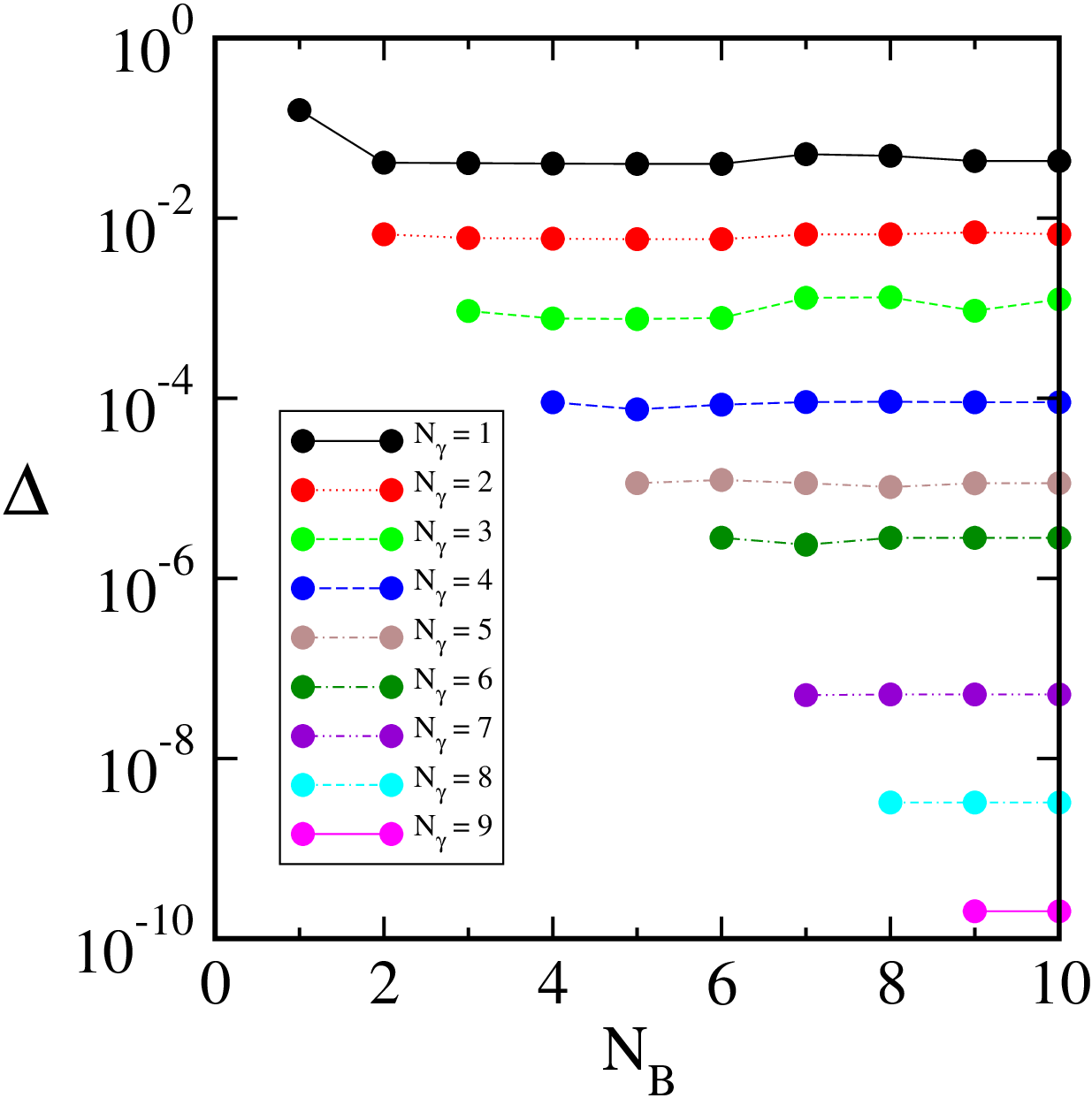}}
\caption{\label{f:chisqr_eos} Curves indicate the values of
  $\Delta(\gamma_k)$ for the $\gamma_k$ that minimize
  $\chi(h_{1c}^i,h_{2c}^i,\gamma_k)$ for different numbers $N_\gamma$
  of spectral parameters, and different numbers $N_B$ of mock binary
  data points.}
\end{figure}

Figure~\ref{f:chisqr_eos} shows the accuracy of the parametric
equations of state whose spectral parameters $\gamma_k$ are set by the
minima of $\chi$ shown in Fig.~\ref{f:chisqr_data}.  These equation of
state errors are measured with the quantity $\Delta$ defined in
Eq.~(\ref{e:Deltadef}).  Like the observational data modeling errors
$\chi$, the equation of state errors $\Delta$ decrease exponentially
as $N_\gamma$ is increased, but are relatively insensitive to $N_B$
for fixed $N_\gamma$.\footnote{The results for $N_\gamma=10$ are not shown in
Figs.~\ref{f:chisqr_data} and \ref{f:chisqr_eos}, because the rates of
convergence decreased abruptly at this point.  This is probably caused
by numerical inaccuracies at the $10^{-10}\sim 10^{-11}$ level in some
part of the code.  Since the source of those errors was not
identified, and since the results for $N_\gamma=10$ appeared to be
unreliable, they were not displayed with the $N_\gamma<10$ results.}

Figures~\ref{f:chisqr_data} and \ref{f:chisqr_eos} show that the
modeling errors $\chi(N_\gamma)$ are comparable to the equation of
state modeling errors $\Delta(N_\gamma)$ for the simple mock data used
in these tests.  This rough comparability of these errors is expected
to apply even for more complicated, more realistic equations of state.
Since representations of more realistic equations of state are
expected to converge more slowly, the modeling errors $\chi$ are also
expected to converge more slowly in those cases.  For smooth equations
of state, these convergence rates are expected to be exponential in
the number of parameters $N_\gamma$.  Equations of state having phase
transitions are expected to converge as a power of $N_\gamma$, with a
power that depends on the order of the phase transition.

%%%%%%%%%%%%%%%%%%%%%%%%%%%%%%%%%%%%%%%%%%%%%%%%%%%%%%%%%%%%%%%%%%%%%%%%%%%%%%%
\section{Discussion}
\label{s:Discussion}

%%%%%%%%%%%%%%%%%%%%%%%%%%%%%%%%%%%%%%%%%%%%%%%%%%%%%%%%%%%%%%%%%%%%%%%%%%%%%%%

The results of the numerical tests in
Sec.~\ref{s:NumericalInversionTests} confirm that the method of
solving the inverse structure problem for neutron-star binaries
outlined in Sec.~\ref{s:Introduction} is mathematically convergent
using data from a randomly chosen ensemble of binaries.  The equation
of state accuracies shown in Fig.~\ref{f:chisqr_eos} are comparable to
the best-fit errors for this equation of state in
Fig.~\ref{f:Delta_n}.  So this method of determining the equation of
state is also very efficient.

Important features of the analysis presented here are its generality
and lack of simplifying assumptions.  No assumptions are made about
the equation of state in the cores of neutron stars other than
thermodynamic stability.  Thermodynamic stability requires the
equation of state function $\epsilon(p)$ to be monotonically
increasing.  It is imposed implicitly by the spectral expansion for
the adiabatic index $\Gamma(h)$ in Eq.~(\ref{e:spectralgamma}) that
ensures $\Gamma(h)\geq 0$.  The analysis here also makes no
simplifying assumptions about the composite deformabilities
$\tilde\Lambda$ of the binaries.  In contrast, the recent analysis of
GW170817 in Ref.~\cite{De2018} assumes the tidal deformabilities of
the two neutron stars are related by $\Lambda_1 M_1^6 = \Lambda_2
M_2^6$, while the analysis in Ref.~\cite{GW170817c} assumes
$\Lambda_2-\Lambda_1$ is a prescribed function of
$\Lambda_1+\Lambda_2$ and the mass ratio $M_2/M_1$.  The analysis here
simply evaluates $\tilde\Lambda$ exactly using
Eq.~(\ref{e:tildeLambda}) in terms of the parametric equation of state
and the central enthalpies of each star.  No additional assumption
about the form of $\tilde\Lambda$ is needed.

The method proposed here for solving the inverse structure problem for
binaries is well posed and admits an exact solution when the number of
data points $N_B$ is greater than or equal to the number of equation
of state parameters $N_\gamma$.  In contrast, the recent analyses in
Refs.~\cite{De2018, Carney2018, GW170817c} attempt to determine four
equation of state parameters using Bayesian statistical methods from
the observation of the single neutron-star binary GW170817.  From the
perspective of the exact problem, it is not possible to determine more
than one equation of state parameter from the observation of a single
binary.  Analyzing a single binary using a four parameter equation of
state model in the exact case could only restrict the four-dimensional
parameter space to some three-dimensional subspace.  To make the
problem well posed, prior constraints on the equation of state
parameters would be needed to fix a particular point on that
three-dimensional parameter subspace.  In the method proposed here for
solving the inverse structure problem, the appropriate dimensional
space of parameters is chosen from the beginning by requiring
$N_\gamma \leq N_B$.  No additional assumptions or prior constraints
on the equation of state parameters are needed.

The ``exact'' equation of state used to create the mock data in these
tests is very simple and very smooth.  Consequently the rate of
convergence of the errors in these tests is probaby faster than it
would be for more realistic equations of state.  The inverse structure
problem for single neutron stars~\cite{Lindblom2014, Lindblom2016} has
been studied using a number of more realistic nuclear-theory model
equations of state. The convergence rates for the equation of state
errors found here are only a bit faster than those found previously
for the smoothest and simplest realistic nuclear-theory based equation
of state models (e.g. PAL6).  Consequently, the expectation is that
the equation of state errors for the binary problem will be similar to
those for the single neutron-star inverse problem studied previously.
A fairly small number of high accuracy measurements from binary
systems should therefore be sufficient to determine the high density
neutron-star equation of state at the fraction of a percent level, if
such high accuracy measurements ever became available.

The mock data used in the analysis in
Sec.~\ref{s:NumericalInversionTests} were constructed with high
precision to allow the mathematical convergence tests of the method to
be confirmed with high confidence.  Those convergence tests were the
primary purpose of this paper.  Observations from real binary systems
will contain significant measurement errors, and those measurement
errors will also contribute to the errors in the equations of state
determined in this way.  More realistic estimates of the equation of
state errors achievable by these methods can only be found therefore
using more realistic mock data for these tests.  The plan for a future
study is to introduce random errors into the mock data with a sequence
of different sizes, e.g. $1\%$, $2\%$, $5\%$, $10\%$, $20\%$, $50\%$
errors, and then to determine how these data errors affect the
inferred equation of state errors.

%%%%%%%%%%%%%%%%%%%%%%%%%%%%%%%%%%%%%%%%%%%%%%%%%%%%%%%%%%%%%%%%%%%%%%%%%%%%%%%
\acknowledgments

I think John Friedman and Massimo Tinto for helpful comments and
suggestions on a draft manuscript of this paper.  This research was
supported in part by NSF grants PHY-1604244 and DMS-1620366.

%%%%%%%%%%%%%%%%%%%%%%%%%%%%%%%%%%%%%%%%%%%%%%%%%%%%%%%%%%%%%%%%%%%%%%%%%%%%%%%%
\vfill\eject

\bibstyle{prd} 
\bibliography{References}

%%%%%%%%%%%%%%%%%%%%%%%%%%%%%%%%%%%%%%%%%%%%%%%%%%%%%%%%%%%%%%%%%%%%%%%%%%%%%%%%
\end{document}